\documentclass[reprint, superscriptaddress, amsmath, amssymb, aps, pra]{revtex4-2}
\usepackage[utf8]{inputenc}
\usepackage[T1]{fontenc}
\usepackage{tabularx}
\usepackage{lineno}
\usepackage{graphicx}
\usepackage{placeins}
\usepackage[dvipsnames]{xcolor}
\usepackage[normalem]{ulem}
\usepackage{xspace}
\usepackage{siunitx}
\usepackage[hypertexnames=false]{hyperref}
\usepackage{orcidlink}
\usepackage{multirow}
\usepackage{booktabs}
\usepackage{siunitx}
\usepackage{cleveref}
\usepackage{natbib}

\newcommand{\refa}{Ref.\xspace}
\newcommand{\qd}{$\mathrm{QD}$\xspace}
\newcommand{\qds}{$\mathrm{QDs}$\xspace}
\newcommand{\qdn}[2]{$\mathrm{QD\;#1}_{#2}$\xspace}
\newcommand{\fidelity}[3]{$#1^{#2}_{#3}$\,\si{\percent}\xspace}
\newcommand{\fidelitywop}[3]{$#1^{#2}_{#3}$\xspace}

\newcommand{\ket}[1]{\left|#1\right>\xspace}

\newcommand{\braketm}[3]{\left<#1|#2|#3\right>\xspace}

\renewcommand\figurename{Fig.}

\makeatletter
\newcommand{\fixref}[1]{\@ifundefined{fixref@#1}{\color{red}INCOMPLETE REFERENCE\color{black}}{\csname fixref@#1\endcsname}}

\expandafter\def\csname fixref@fig:figure1\endcsname{Fig. \hyperref[fig:figure1]{1}}
\expandafter\def\csname fixref@fig:figure2\endcsname{Fig. \hyperref[fig:figure2]{2}}
\expandafter\def\csname fixref@fig:figure3\endcsname{Fig. \hyperref[fig:figure3]{3}}
\expandafter\def\csname fixref@fig:figure4\endcsname{Fig. \hyperref[fig:figure4]{4}}

\expandafter\def\csname fixref@tab:table1\endcsname{Table \hyperref[tab:table1]{1}}

\expandafter\def\csname fixref@ext:fig:extended_data_figure_1\endcsname{Extended Data Fig. \hyperref[ext:fig:extended_data_figure_1]{1}}
\expandafter\def\csname fixref@ext:fig:extended_data_figure_2\endcsname{Extended Data Fig. \hyperref[ext:fig:extended_data_figure_2]{2}}
\expandafter\def\csname fixref@ext:fig:extended_data_figure_3\endcsname{Extended Data Fig. \hyperref[ext:fig:extended_data_figure_3]{3}}
\expandafter\def\csname fixref@ext:fig:extended_data_figure_4\endcsname{Extended Data Fig. \hyperref[ext:fig:extended_data_figure_4]{4}}
\expandafter\def\csname fixref@ext:fig:extended_data_figure_5\endcsname{Extended Data Fig. \hyperref[ext:fig:extended_data_figure_5]{5}}
\expandafter\def\csname fixref@ext:fig:extended_data_figure_6\endcsname{Extended Data Fig. \hyperref[ext:fig:extended_data_figure_6]{6}}

\expandafter\def\csname fixref@sup:note:supplementary_note_1\endcsname{Supplementary Note I}
\expandafter\def\csname fixref@sup:fig:supplementary_figure_1\endcsname{Supplementary Figure \hyperref[sup:fig:supplementary_figure_1]{1}}
\expandafter\def\csname fixref@sup:vid:video1\endcsname{Supplementary Video 1}
\expandafter\def\csname fixref@sup:vid:video2\endcsname{Supplementary Video 2}
\makeatother

\begin{document}
\title{Conveyor-mode electron shuttling through a T-junction in Si/SiGe}

\author{Max Beer \orcidlink{0009-0007-9002-256X}}
\affiliation{JARA-FIT Institute for Quantum Information, Forschungszentrum Jülich GmbH and RWTH Aachen University, 52074 Aachen, Germany}
\author{Ran Xue}
\affiliation{JARA-FIT Institute for Quantum Information, Forschungszentrum Jülich GmbH and RWTH Aachen University, 52074 Aachen, Germany}
\author{Lennart Deda}
\affiliation{JARA-FIT Institute for Quantum Information, Forschungszentrum Jülich GmbH and RWTH Aachen University, 52074 Aachen, Germany}
\author{Stefan Trellenkamp}
\affiliation{Helmholtz Nano Facility (HNF), Forschungszentrum Jülich, Jülich, Germany}
\author{Jhih-Sian Tu}
\affiliation{Helmholtz Nano Facility (HNF), Forschungszentrum Jülich, Jülich, Germany}
\author{Paul Surrey}
\affiliation{JARA-FIT Institute for Quantum Information, Forschungszentrum Jülich GmbH and RWTH Aachen University, 52074 Aachen, Germany}
\author{Inga Seidler}
\affiliation{JARA-FIT Institute for Quantum Information, Forschungszentrum Jülich GmbH and RWTH Aachen University, 52074 Aachen, Germany}
\author{Hendrik Bluhm}
\affiliation{JARA-FIT Institute for Quantum Information, Forschungszentrum Jülich GmbH and RWTH Aachen University, 52074 Aachen, Germany}
\affiliation{ARQUE Systems GmbH, 52074 Aachen, Germany}
\author{Lars R. Schreiber \orcidlink{0000-0003-0904-9612}}
\email{lars.schreiber@physik.rwth-aachen.de}
\affiliation{JARA-FIT Institute for Quantum Information, Forschungszentrum Jülich GmbH and RWTH Aachen University, 52074 Aachen, Germany}
\affiliation{ARQUE Systems GmbH, 52074 Aachen, Germany}

\begin{abstract} 
Conveyor-mode shuttling in gated Si/SiGe devices enables adiabatic transfer of single electrons, electron patterns and spin qubits confined in quantum dots across several microns with a scalable number of signal lines. To realize their full potential, linear shuttle lanes must connect into a two-dimensional grid with controllable routing. We introduce a T-junction device linking two independently driven shuttle lanes. Electron routing across the junction requires no extra control lines beyond the four channels per conveyor belt. We measure an inter-lane charge transfer fidelity of $F =$ \fidelity{100.0000000}{+0}{-9\times 10^{-7}} at an instantaneous electron velocity of \SI{270}{\milli\meter\per\second}. The filling of 54 quantum dots is controlled by simple atomic pulses, allowing us to swap electron patterns, laying the groundwork for a native spin-qubit SWAP gate. This T-junction establishes a path towards scalable, two-dimensional quantum computing architectures with flexible spin qubit routing for quantum error correction.
\end{abstract}

\flushbottom
\maketitle

\thispagestyle{empty}

Electron and hole spin quantum computing in electrostatically defined SiGe quantum dots (\qds) have progressed considerably \cite{burkardSemiconductorSpinQubits2023}: Single- and two-qubit gate fidelities exceeding the error correction threshold \cite{noiriFastUniversalQuantum2022,unseldBasebandControlSingleelectron2025,millsTwoqubitSiliconQuantum2022,xueQuantumLogicSpin2022} as well as scaling of multi-qubit devices up to twelve qubits \cite{george12SpinQubitArraysFabricated2025} have been demonstrated.
These scaling efforts manifest as increasingly large (bi-) linear \cite{philipsUniversalControlSixqubit2022, weinsteinUniversalLogicEncoded2023, zhangUniversalControlFour2025, george12SpinQubitArraysFabricated2025} and dense, two-dimensional \qd arrays \cite{hendrickxSweetspotOperationGermanium2024, johnRobustLocalisedControl2025}.
Such dense architectures as well as large linear arrays suffer from limited scalability due to wiring bottlenecks and difficulties in implementing the long-range qubit coupling required \cite{vandersypenInterfacingSpinQubits2017, boterSpiderwebArraySparse2022}, while two-dimensional qubit topologies with millions of physical qubits are needed for the implementation of many efficient quantum error correction algorithms \cite{taylorFaulttolerantArchitectureQuantum2005, ginzelScalableParityArchitecture2024a}. 

To solve these challenges, qubit shuttling has been investigated intensely.
The two dominant methods of qubit shuttling in \qds are bucket-brigade \cite{fujitaCoherentShuttleElectronspin2017,zwerverShuttlingElectronSpin2023,fosterDephasingErrorDynamics2025,desmetHighfidelitySinglespinShuttling2025} and conveyor-mode \cite{seidlerConveyormodeSingleelectronShuttling2022,xueSiSiGeQuBus2024,struckSpinEPRpairSeparationConveyormode2024,volmerMappingValleySplitting2024,desmetHighfidelitySinglespinShuttling2025} shuttling. The latter features higher spin shuttle fidelity and has been demonstrated as the more scalable solution \cite{desmetHighfidelitySinglespinShuttling2025, xueSiSiGeQuBus2024, struckSpinEPRpairSeparationConveyormode2024, volmerReductionImpactLocal2025, langheinrichFabricationSingleelectronShuttling2025} as it demands only a small and constant number of control lines and signals independent of the shuttling distance \cite{langrockBlueprintScalableSpin2023, kunneSpinBusArchitectureScaling2024,davidLongDistanceSpin2024}. However, the required extension of conveyor-mode shuttling of qubits into two dimensions has not been demonstrated so far. Shuttle junctions and qubit routing pose a challenge even for other, arguably more mature qubit platforms \cite{jainPenningMicrotrapQuantum2024,barredoAtombyatomAssemblerDefectfree2016, delaneyScalableMultispeciesIon2024}.  Surface-code quantum error correction for spin-qubits requires thousands of physical qubits to encode one logical qubit thus millions of single electron spins need to be positioned and controlled. This sets high demands on the fidelity for controlling these electrons, while control signals must be simple to co-integrate signal generation and routing on the chip \cite{boterSpiderwebArraySparse2022, kunneSpinBusArchitectureScaling2024}.    

Here we demonstrate a key building block, a two-dimensional intersection of orthogonal one-dimensional conveyor-belt shuttle lanes. This T-junction consists of 54 \qds and enables flexible routing of single electrons.
Despite its gate design complexity, all routing and filling of all \qds is controlled by only eight simple voltage signals. We observe single electron transport across the \SI{10}{\micro\meter} $\times$ \SI{5}{\micro\meter} device with an inter-lane charge transfer fidelity of $F =$ \fidelity{100.0000000}{+0}{-9\times 10^{-7}} at an instantaneous electron velocity of \SI{270}{\milli\meter\per\second}. We demonstrate the first step towards a spin-SWAP gate native to our device topology by permuting initialized \qd fillings also referred to as electron patterns \cite{xueSiSiGeQuBus2024}.

\section*{T-junction device and atomic pulses}
\begin{figure*}
    \centering
    \includegraphics[scale=1]{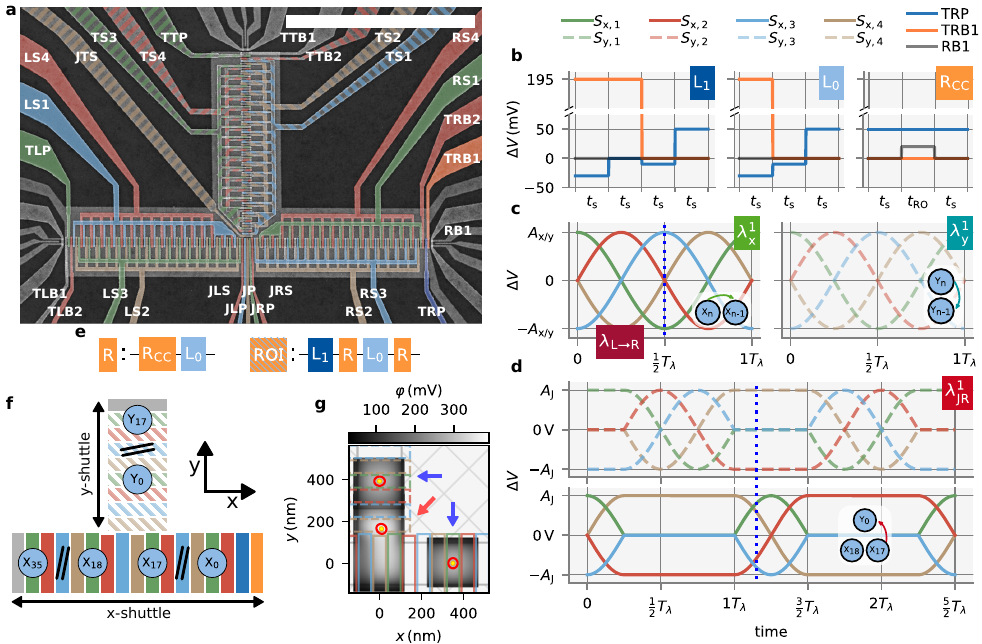}
    \caption{\textbf{T-junction device and atomic pulses.} \textbf{a} False color scanning electron micrograph of the surface of a device nominally identical to the device measured, including gate labels. Gate color coding is consistent between all sub-figures and indicates gates comprising gate sets $\mathrm{S}_{\mathrm{x}, i}$ and $\mathrm{S}_{\mathrm{y}, i}$ (see methods). The scale bar is \SI{5}{\micro\meter}. \textbf{b} Atomic pulses for charge initialization and readout. The indicated voltages are typical values. Typical values are $t_{\mathrm{S}} \approx $ \SI{10}{\milli\second} and $t_{\mathrm{RO}} \approx $ \SI{100}{\milli\second}. \textbf{c} Horizontal and vertical charge shuttle atomic pulses. The first half of $\lambda_x^1$, up to the dashed line, describes $\lambda_{\mathrm{L}\rightarrow\mathrm{R}}$. \textbf{d} Junction transfer atomic pulse. Pulses for finger gates in x- and y-shuttle are shown on the bottom and top, respectively. Note that the x- and y-shuttle are not operated simultaneously. \textbf{e} Commonly used composite pulses. R: destructive readout pulse. ROI: readout and initialization calibration pulse. \textbf{f} Nominal dot positions in the idle state. In the readout state (see methods), all \qdn{X}{n} are shifted two gates towards the right. Periodic continuation of more \qds are omitted but indicated with (//) symbols. \textbf{g} Simulated potentials $\varphi$ and electron ground state probability densities (yellow-red circular shapes) at three positions in the device corresponding to three points in time. Claviature gates are shown as outlines and screening gates are indicated as a light gray grid. Blue arrows: Electron states in \qdn{X}{16} and \qdn{Y}{0} in the idle state. Red arrow: Electron during inter-lane junction transfer, time indicated by blue dashed line in (\textbf{d}). \SI{95}{\percent}, \SI{68}{\percent}, \SI{1}{\percent} probability density levels shown in yellow, orange and red, respectively. (time-dependence of simulations in Supplementary Movies 1 and 2)}
    \label{fig:figure1}
\end{figure*}

The T-junction is composed of two individual QuBus shuttle lanes, labeled x-shuttle and y-shuttle, joined at a $90^\circ$ angle with a single electron transistor (SET) formed at the exterior-facing end of each lane (\fixref{fig:figure1}a). Electrostatic gates used in the operation of the device are labeled in \fixref{fig:figure1}a. The large number of gates and device terminals allows for more flexibility and fine control over the electron confinement potential than actually required to operate the device in this work. For this reason, gates are grouped in connected gate sets $\mathrm{S}_{j, i}$ ($i\in\{1,2,3,4\}$, $j\in\{\mathrm{x}, \mathrm{y}\}$) as indicated by color in \fixref{fig:figure1}a (see methods for details). The same voltage is applied to each gate in a set, significantly reducing the effective number of terminals required to operate the T-junction.

We define a pulse $P$ as time-dependent voltage sequences simultaneously applied to a few gate sets (in addition to a constant DC voltage, details on notation in methods). A pulse that is not composed of other pulses is called an atomic pulse, of which we define eight (1 and 0 electron loading $\mathrm{L}_1$, $\mathrm{L}_0$; charge readout $\mathrm{R}_\mathrm{CC}$ and preparation $\lambda_\mathrm{L\rightarrow R}$ (\fixref{fig:figure1}b); linear charge shuttling $\lambda^1_\mathrm{x}$, $\lambda^1_\mathrm{y}$ (\fixref{fig:figure1}c); T-junction transfers $\lambda^1_\mathrm{JR}$, $\lambda^1_\mathrm{JL}$ (\fixref{fig:figure1}d)), each achieving a specific function explained below or in methods. We abbreviate often used pulses related to initialization and readout by $\mathrm{R}$ and $\mathrm{ROI}$ (\fixref{fig:figure1}e). In total, only eleven voltage channels are required to apply all atomic pulses. Despite the device complexity, these eight atomic pulses are sufficient to operate the entire device.

Typical for the conveyor-belt shuttling approach \cite{seidlerConveyormodeSingleelectronShuttling2022, langrockBlueprintScalableSpin2023, xueSiSiGeQuBus2024, kunneSpinBusArchitectureScaling2024}, \qds are formed under every fourth gate in the x- and y-shuttle lane, here in total 54 – labeled \qdn{X}{i} and \qdn{Y}{i} (\fixref{fig:figure1}f). Forward conveyor-belt shuttling in the x-(y-)shuttle lane means a continuous transfer of a \qd positioned at \qdn{X}{n} to \qdn{X}{n-1} (\qdn{Y}{n} to \qdn{Y}{n-1}).
To facilitate forward conveyor-belt shuttling, shuttle gates in the x- and y-shuttle are pulsed by the voltage 
\begin{gather*}
    \Delta V_{g\,\in\,\mathrm{S}_{j, i}}\left(t\right) = A_j\sin\left(-\frac{2\pi t}{T_{\lambda}} + i\cdot \frac{\pi}{2}\right) \hspace{1.5em} \begin{array}{l}i\in\{1, 2, 3, 4\}\\j\in\{\mathrm{x}, \mathrm{y}\}\end{array}
\end{gather*}
where $T_\lambda$ is the shuttle period and $A_\mathrm{x}$, $A_\mathrm{y}$ are the drive amplitudes for the x- and y-shuttle, respectively. The voltage pulses applied to $\mathrm{S}_\mathrm{x, 3}$ are also applied to TRP.
A single period $T_\lambda$ of this sequence, applied to either the x- or y-shuttle, defines atomic pulses $\lambda^1_\mathrm{x}$ or $\lambda^1_\mathrm{y}$ (\fixref{fig:figure1}c), continuously moving a \qd a distance of four gates, i.e. $\lambda =$ \SI{280}{\nano\meter} with an instantaneous shuttle velocity $v_\lambda = \lambda/T_{\lambda}$. 

The atomic pulse $\lambda^1_\mathrm{JR}$ is specific to the T-junction and transfers the charge occupancy from \qdn{X}{17} continuously into \qdn{Y}{0} (finite-element simulation in \fixref{fig:figure1}g with details in Supplementary Note I). The transfer is realized by coordinated conveyor-mode shuttling with a drive amplitude $A_\mathrm{J}$ and velocity $v_\mathrm{J}$ in the x- and y-shuttle. It includes halting and reversing both shuttle lanes appropriately at $T_\lambda/4$ intervals (\fixref{fig:figure1}d), but otherwise signal bandwidth requirements are similar to $\lambda^1_\mathrm{x}$ and $\lambda^1_\mathrm{y}$. The analogue atomic pulse transferring the charge occupancy directly from \qdn{X}{18} into \qdn{Y}{0}, $\lambda^1_\mathrm{JL}$, is additionally defined, differing only by the distance and direction of conveyor-mode shuttling in the x-shuttle (\fixref{ext:fig:extended_data_figure_1}). 

\section*{Single electron shuttling} 
\begin{figure*}
    \centering
    \includegraphics[scale=1]{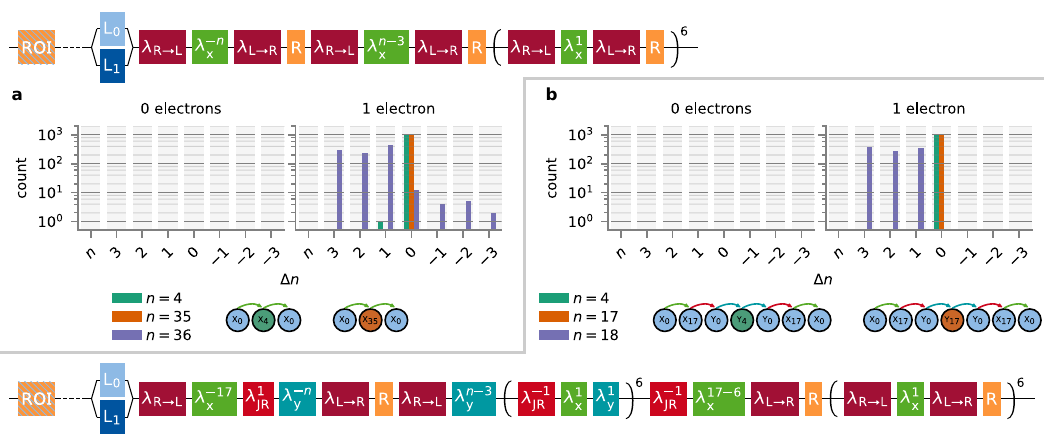}
    \caption{\textbf{Verifying shuttling in the x- and y-shuttles.} Applied pulses and readout histograms with one and zero electrons loaded, the latter as reference pulse. Branching notation in the pulses indicate that the whole pulse is repeated once for each branch, with the exception of the ROI-pulse separated by dashed lines, which is only performed for the first branch repetition. Histograms show the sum of \qdn{X}{0} occupancy readout counts separated for the pulse branches with either one electron or zero electrons loaded. The \qds along the shuttle paths are shown next to the legends. $n_\mathrm{total} = 1000$, $A_\mathrm{x} = A_\mathrm{y} = A_\mathrm{J} =$ \SI{260}{\milli\volt} and $v_\lambda = v_\mathrm{J} =$ \SI{270}{\milli\meter\per\second} for all measurements. Results for $v_\lambda = v_\mathrm{J} =$ \SI{28}{\milli\meter\per\second} shown in \fixref{ext:fig:extended_data_figure_2}. \textbf{a} Shuttling in the x-shuttle. \textbf{b} Shuttling in the x-shuttle up to \qdn{X}{17} followed by y-shuttle.}
    \label{fig:figure2}
\end{figure*}

\begin{figure}
    \centering
    \includegraphics[scale=1]{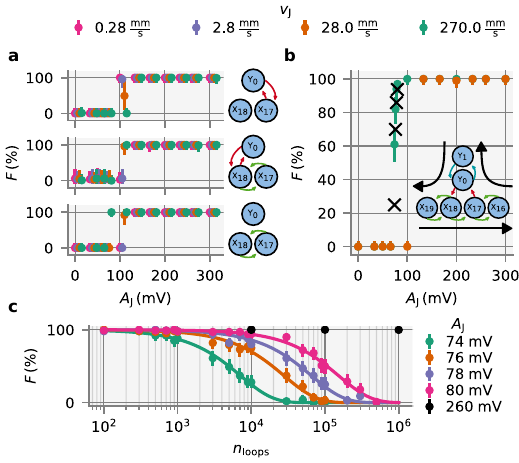}
    \caption{\textbf{Parameter variation and charge looping within the T-junction.}
    $n_\mathrm{total} = 100$, $A_\mathrm{x} = A_\mathrm{y} = $ \SI{260}{\milli\volt} and $v_\lambda = $ \SI{270}{\milli\meter\per\second} for all measurements if not otherwise indicated. \textbf{a} Pulse fidelity at different drive amplitudes $A_\mathrm{J}$ and shuttle velocities $v_\mathrm{J}$ during shuttling in the junction. Additional shuttling into the y-shuttle is not shown schematically. Data points offset in $A_\mathrm{J}$ by \SI{5}{\milli\volt} per $v_\mathrm{J}$ value starting from $v_\mathrm{J} =$ $0.28\,$\si{\milli\meter\per\second}. Error bars are $4\sigma^\pm_{F_\mathrm{total}}$. \textbf{b} Fidelity at different drive amplitudes $A_\mathrm{J}$ and shuttle velocities $v_\mathrm{J}$ during looped inter-lane junction transfer for $n_\mathrm{loops} = 10,000$. Black crosses are interpolated from $F_1$ extracted in (\textbf{c}). Error bars are $1\sigma^\pm_{F_\mathrm{total}}$. \textbf{c} Pulse fidelity for different numbers of loops $n_\mathrm{loops}$ at $v_\mathrm{J}=$ \SI{270}{\milli\meter\per\second} and select drive amplitudes $A_\mathrm{J}$ for the same pulse as (\textbf{b}). $n_\mathrm{total} = 1000$ and $n_\mathrm{total} = 490$ for $A_\mathrm{J} = $ \SI{260}{\milli\volt} at $n_\mathrm{loops} = 10^5$ and $n_\mathrm{loops} = 10^6$, respectively. Error bars are $1\sigma^\pm_{F_\mathrm{total}}$. }
    \label{fig:figure3}
\end{figure}

This work focuses on shuttling and routing across a T-junction.
Because no charge sensor is implemented directly at the densely gated center of the T-junction, we investigate the initial success of routing across the T-junction by continuing to linearly shuttle in the x-(y-)shuttle, until the shuttled charge encounters the left (upper) end of the x-(y-)shuttle lane, which can be detected \cite{xueSiSiGeQuBus2024}.
Then, we amplify the inter-lane junction transfer infidelity by repeatedly looping around the T-junction with a single electron.

First we verify shuttling in the x-shuttle lane and thus shuttle straight through the T-junction.
The pulse (\fixref{fig:figure2}a with $A_\mathrm{x}  = 260\,\mathrm{mV}$ and $v_\lambda = $ \SI{270}{\milli\meter\per\second}) shuttles the initialized charge from \qdn{X}{0} into \qdn{X}{n}, followed by readout of \qdn{X}{0} to verify that the charge was shuttled away and finally returns it to \qdn{X}{0}, shuttling for three periods more than required to return the charge. With in total seven readouts (three shuttling periods each before and after the charge is expected to return), we detect unexpectedly advanced or delayed charge. These readouts are labeled in the resulting data by their $\Delta n$ value, which is $\Delta n = n$ for the first readout, when the charge is expected to be farthest from the SET and $\Delta n \in [-3, 3]$, indicating the readouts relative to the expected $\Delta n = 0$ readout ($\Delta n > 0$ and $\Delta n < 0$ corresponding to advanced or delayed charge, respectively). This pulse is applied with zero electrons loaded as a reference and then, immediately afterwards, with one electron to ensure the measured electron occupancy is only a consequence of the deliberate initialization of \qdn{X}{0}.
A shuttle event is called successful, if the charge is detected only at $\Delta n=0$ and no charge is detected during the reference sequence.
By a modified pulse (\fixref{fig:figure2}b), we verify turning within the T-junction by inter-lane junction transfer ($v_\mathrm{J} = v_\lambda$, $A_\mathrm{J}= A_\mathrm{x}= A_\mathrm{y}$), shuttling until \qdn{Y}{n}.

When initializing \qdn{X}{0} with one electron, that electron is observed as expected in readout $\Delta n=0$ when attempting to shuttle for $n\in\{4, 35\}$ ($n\in\{4, 17\}$ for turning) during almost all repetitions (\fixref{fig:figure2}a,b).
No electrons are detected if zero electrons are loaded, as expected. For $n=36$ ($n=18$ for turning), far fewer electrons are detected in readout $\Delta n = 0$, while many electrons are detected earlier in readouts $\Delta n \in \{1,2,3\}$.
This is our fingerprint that the electron actually reaches the end of the x-(y-)shuttle lane at which a large barrier is formed by gate TLB1 (TTB1) scattering the electron across neighboring \qds.
Hence, it demonstrates that we routed a single charge across the T-junction. 
We evaluate the pulse fidelity $F_\mathrm{total}$ as defined in methods (results noted in \fixref{tab:table1}).
The charge control in the T-junction is reliable across 1000 repetitions and independent of shuttled distance – we only observe one unsuccessful shuttle event for $v_\lambda = $ \SI{28}{\milli\meter\per\second} and \SI{270}{\milli\meter\per\second}.
A deviation of $F_\mathrm{total}$ from \SI{100}{\percent} could not be determined as the duration of the pulses is limited by initialization and readout ($\mathrm{L}_0$, $\mathrm{L}_1$ and $\mathrm{R}_\mathrm{CC}$).

When shuttling in the junction, three main shuttle paths exist, each with its own range of possible drive amplitudes and shuttle velocities.
To explore the optimal parameter $(A_{\mathrm{J}}, v_{\mathrm{J}})$ space for operation of the T-junction, we apply several pulses that first shuttle the initialized \qd close to the junction using the good drive amplitudes $A_\mathrm{x} = A_\mathrm{y} =$ \SI{260}{\milli\volt} and shuttle velocity $v_\lambda =$ \SI{270}{\milli\meter\per\second} and then either route the initialized \qd into the y-shuttle lane using $\lambda^{1}_{\mathrm{JR}}$ or $\lambda^{1}_{\mathrm{JL}}$ or shuttle from \qdn{X}{16} to \qdn{X}{18} using $\lambda^{-2}_{\mathrm{x}}$, while varying the drive amplitude $A_{\mathrm{J}}$ at shuttle velocity $v_{\mathrm{J}}$ (\fixref{fig:figure3}a).
Note that we set $A_{\mathrm{x}} = A_{\mathrm{J}}$ and $v_\lambda = v_{\mathrm{J}}$ for the final path when shuttling from \qdn{X}{16} to \qdn{X}{18} only.
The full pulses used here are shown in \fixref{ext:fig:extended_data_figure_3}.
The pulse fidelities observed remain close to \SI{100}{\percent} for $A_{\mathrm{J}} \geq $  \SI{133}{\milli\volt} for $\lambda^{1}_{\mathrm{JR}}$, $\lambda^{1}_{\mathrm{JL}}$ and $\lambda^{-2}_{\mathrm{x}}$, with only a minor dependence of the minimum reliable drive amplitude on the instantaneous shuttle velocity during junction transfer shuttling $v_\mathrm{J}$.
While the cable bandwidth limits the investigated $v_\mathrm{J}$ to one order of magnitude lower than required for spin-coherent shuttling \cite{langrockBlueprintScalableSpin2023}, we observe no significant reduction of pulse fidelity up to the maximum possible $v_\mathrm{J}$.
The drive amplitudes allowing for successful inter-lane junction transfer shuttling are similar to the required drive amplitude for linear shuttling in \refa\cite{xueSiSiGeQuBus2024}, performed on a different device design, albeit on a nominally identical heterostructure.

\section*{Charge looping within the T-junction}
As the previous pulses transfer the initialized charge through the junction only two times per pulse, we expect their fidelity to be dominated by charge state preparation and readout (SPAM) errors and possibly by errors occurring when shuttling along the x- and y-shuttle lanes. Additionally, charge initialization and readout contribute the majority of measurement time, thereby limiting the total number of repetitions.
Hence, these pulses are not able to resolve the small infidelity of inter-lane junction transfer ($\lambda^{\pm1}_{\mathrm{JR}}$ and $\lambda^{\pm1}_{\mathrm{JL}}$) at the center of the T-junction.
We therefore modify the pulse such that the initialized charge is instead shuttled into \qdn{X}{16} using again the known good drive amplitudes $A_\mathrm{x} = A_\mathrm{y} =$ \SI{260}{\milli\volt}.
Then, we loop the charge $n_{\mathrm{loops}}$-times within the T-junction, following the looped shuttle path indicated in \fixref{fig:figure3}b (inset) before returning to \qdn{X}{0}, performing a readout procedure identical to the previous pulses (results noted in \fixref{fig:figure3}b, full pulse shown in \fixref{ext:fig:extended_data_figure_3}).

During each loop, the nominally empty shuttle lane which the electron is expected to have left due to the applied junction transfer atomic pulse ($\lambda^{1}_{\mathrm{JR}}$ or $\lambda^{-1}_{\mathrm{JL}}$) is operated for $25$ periods ($\lambda_{\mathrm{x}}^{25}$ or $\lambda_{\mathrm{y}}^{-25}$).
This large number of additional shuttle pulses provides us high confidence that the electron is shuttled away from the center of the T-junction if the junction transfer atomic pulse fails.
Thus, the detection of a failed $\lambda^{1}_{\mathrm{JR}}$ or $\lambda^{-1}_{\mathrm{JL}}$ atomic pulse is ensured. 
As $n_{\mathrm{loops}}$ increases, the ratio of electron loading events and readouts (i.e. SPAM) against the number of shuttle atomic pulses decreases, rendering the resulting fidelity more sensitive to shuttle errors occurring during looping within the T-junction. This strategy allows us to measure the shuttle fidelity of looping within T-junction with extremely high accuracy.

The fidelity per charge loop $F_1$ (see methods) does not significantly drop below \SI{100}{\percent} for junction transfer drive amplitudes $A_\mathrm{J} \geq$ \SI{133}{\milli\volt}. The charge loops are repeated for select $A_\mathrm{J}$ at $v_\mathrm{J} =$ \SI{270}{\milli\meter\per\second} as a function of $n_\mathrm{loops}$  and we fit $F_\mathrm{total}(n_\mathrm{loops})=F_1^{{n_\mathrm{loops}}}$ to the resulting data (\fixref{fig:figure3}c, fits indicated as solid lines).
The data follow these fits well, indicating statistical independence of failure of junction transfer per loop.
As a sanity check, we calculate $F_\mathrm{total}(A_\mathrm{J}, n_\mathrm{loops}=10,000)$ from the determined $F_1(A_\mathrm{J})$ (crosses in \fixref{fig:figure3}b), which agrees with the directly measured results (dots in \fixref{fig:figure3}b) at $v_\mathrm{J} = $ \SI{270}{\milli\meter\per\second}.

To further resolve the transfer fidelity, the experiment is repeated at $A_{\mathrm{J}} =$ \SI{260}{\milli\volt} for large $n_\mathrm{loops}$ with complete results noted in \fixref{tab:table1}.
We additionally reduce the junction transfer drive amplitude to its approximate reliable minimum $A_{\mathrm{J}} = $ \SI{100}{\milli\volt}, which at $n_\mathrm{loops}=10^6$ and $v_\mathrm{J}$ = \SI{270}{\milli\meter\per\second} yields $F_1 = $ \fidelity{100.0000000}{+0}{-9\times 10^{-7}}, not significantly differing from the result at $A_{\mathrm{J}} = $ \SI{260}{\milli\volt}.

We can feed the applied pulses back to the simulation to investigate the expected quantum states during shuttling with $A_\mathrm{J} =$ \SI{260}{\milli\volt} and $A_\mathrm{J} =$ \SI{100}{\milli\volt} (Supplementary Note I).
The simulation indicates that the electron is continuously moved around the T-junction.

\section*{Multiple electron operation}
\begin{figure*}
    \centering
    \includegraphics[scale=1]{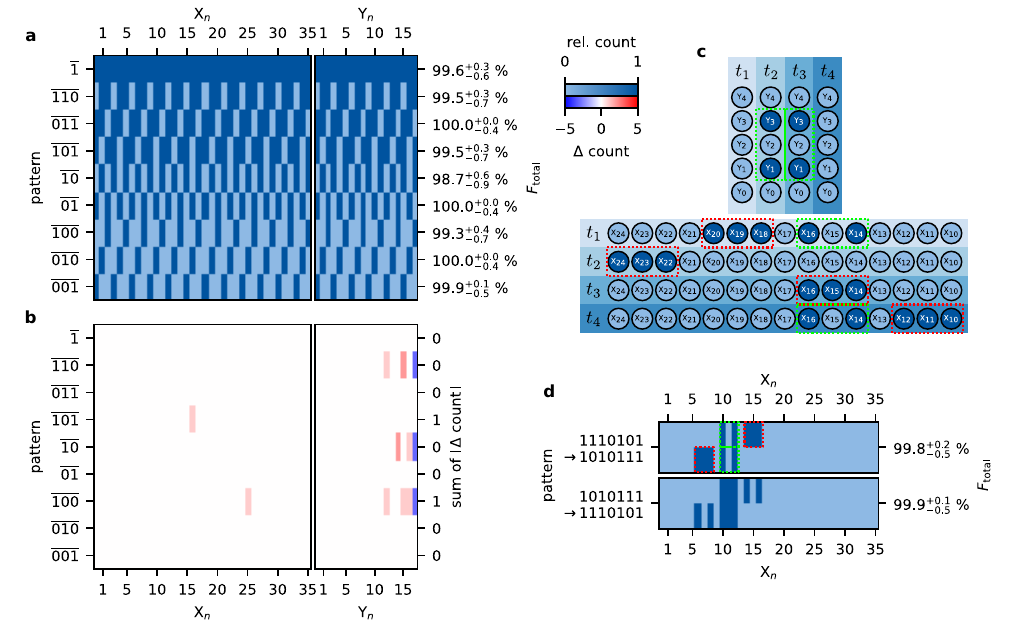}
    \caption{\textbf{Multiple electron operation.} Color bars apply to (\textbf{a}), (\textbf{b}) and (\textbf{d}). Blue color bar: average \qd occupancy over pulse repetitions. Blue-red color bar: Sum of \qd occupancy deviating from expectation over pulse repetitions. $n_\mathrm{total} = 1000$ for each electron pattern and SWAP. If two electrons are detected in the same \qd, the total result of the pulse is excluded from occupancy counts while remaining included in the calculation of $F_\mathrm{total}$. \textbf{a} Results for loading periodic electron occupancy patterns into the x- and y-shuttle. The patterns are held in both channels simultaneously. Overline notation to the left of the plot identifies the periodic patterns. The total fidelity of each pattern is displayed to the right. \textbf{b} Deviation from expectation in (\textbf{a}). Sum of absolute sum of deviations over all electron counts in each repetition over all repetitions is noted to the right. \textbf{c} Expected charge configuration of a section of the x- and y-shuttle when performing an electron pattern SWAP. Dark blue \qd: one electron in \qd. Light blue \qd: no electron in \qd. The leading and following patterns are outlined in red and green, respectively. Labels $t_1$ through $t_4$ indicate four time steps during the pulse. $t_1$: The two sequential patterns are shuttled to the junction. $t_2$: The following pattern is transferred into the y-shuttle. $t_3$: The leading pattern is retracted. $t_4$: The following pattern is retrieved and the pattern SWAP is complete. \textbf{d} Results for pattern SWAP. Patterns are highlighted matching (\textbf{c}). Pulse fidelities are shown to the right. The deviation from expectation is noted in \fixref{ext:fig:extended_data_figure_4}.}
    \label{fig:figure4}
\end{figure*}

Next, we verify whether arbitrary, periodic electron patterns – the filling (empty or singly occupied) of multiple \qds in the device \cite{xueSiSiGeQuBus2024} – can be simultaneously prepared in both the x- and the y-shuttle.
We start by loading different electron patterns into the right half of the x-shuttle, followed by shuttling the pattern via repeated $\lambda^1_{\mathrm{JR}}\lambda^{-1}_{\mathrm{x}}\lambda^{-1}_{\mathrm{y}}$-pulses into the y-shuttle and loading a new pattern into the entire x-shuttle, all with $A_\mathrm{x} = A_\mathrm{y} = A_\mathrm{J} = $ \SI{260}{\milli\volt} and $v_\lambda = v_\mathrm{J} = $ \SI{270}{\milli\meter\per\second}. Finally, we read-out both shuttle lanes in reverse order (\fixref{fig:figure4}a, exemplary pulse in \fixref{ext:fig:extended_data_figure_5}c).
Note that in particular for patterns with high \qd occupancy, e.g. $\overline{1}$, it is impossible to differentiate unintentional electron tunneling from a SPAM error during $\mathrm{L}_1$ resulting in the detection of two electrons in the same \qd.
Therefore, we eliminate all pulses with two electrons detected in any \qd from the \qd occupancy counts.
In the calculation of $F_\mathrm{total}$, we count a pulse as successful (failed) if one or two electrons are detected in a \qd with one electron (zero electrons) expected (see methods).

Remarkably, simultaneous shuttling and routing of up to 54 electrons in the two independent shuttle lanes  yields fidelities of $F_\mathrm{total}\approx$ \SI{99}{\percent}. The main error is detection of additional electrons (red in \fixref{fig:figure4}b) due to unintentional loading of electrons (i.e. SPAM errors) caused by uncompensated drift in the $\mathrm{L}_0$ atomic pulse. Importantly, tunneling across neighboring \qds is not observed except from \qdn{Y}{17} (red/blue in \fixref{fig:figure4}b), presumably due to the large potential barrier at the edge of the y-shuttle.

An interesting feature of the T-junction is the ability to perform a native spin-SWAP gate by swapping electron positions. This can replace the exchange-based spin-SWAP gate, the fidelity of which is low due to need of diabatic switching of the exchange interaction \cite{nowackSingleShotCorrelationsTwoQubit2011}.
Here, the SWAP operation is emulated without spin control by swapping the electron order of two consecutively prepared electron patterns utilizing the independent control of the x- and y-shuttle (\fixref{fig:figure4}c with the pulse shown in \fixref{ext:fig:extended_data_figure_5}d).
We observe pulse fidelities $F_\mathrm{total} \geq $ \SI{99.8}{\percent} (\fixref{fig:figure4}d), independent of the initial order of the two prepared patterns.

\section*{Conclusions}
We demonstrate single electron routing across a conveyor-mode shuttling based T-junction in Si/SiGe. In general, the charge shuttle fidelity is high with particularly high charge transfer fidelity of $F =$ \fidelity{100.0000000}{+0}{-9\times 10^{-7}} in the center of the T-junction.
This is remarkable as most gates of the T-junction are connected and no tuning of voltages applied to individual gates is performed. Such fine-tuning is common for state-of-the-art qubit devices, posing a challenge for device scalability.
We operate the T-junction device that consists of 54 \qds by only eleven voltage channels, three of which required solely for charge initialization and readout by a single-electron transistor (SET). In total, we need only 5 DC voltages including one voltage to form the barriers for the top and left end of the shuttle lanes and 4 voltages to form the SET.
Eight short atomic voltage pulses, from which we easily compile complex pulses by an arbitrary waveform generator, are sufficient to operate the whole T-junction device.
These pulses can control up to 54 electrons with arbitrary single electron fillings of all \qds with high accuracy.
In particular, we use our T-junction to swap electron positions with fidelity exceeding the fidelity of an exchange-based SWAP gate commonly used for spin qubits.

Controlling positions of many single electrons with simple voltage pulses in two dimensions has important implications especially addressing the scalability of spin-based quantum computation.
The voltage tuning is simple and fast and fewer signal generators are needed.
In fact, our T-junction has been operated in two thermal cycles without need of voltage fine-tuning.
Such high level of control can be exploited for novel low-power classical electronics operating with single electrons in \qds below \SI{1}{\kelvin}.
It is also promising for fault-tolerant quantum computing with spin qubits, for which millions of electrons have to be positioned in a two-dimensional matrix of shuttle lanes.
In such lanes, spin-qubit shuttling with high velocity and spin manipulation has been demonstrated.
Ultimately, in a sufficiently sparse two-dimensional qubit architecture, classical control electronics could be co-integrated on the quantum chip \cite{kunneSpinBusArchitectureScaling2024}.

\FloatBarrier

\clearpage
\section*{Methods}
\subsection*{Device fabrication}
The device is fabricated on an undoped $\mathrm{Si}_{0.70}\mathrm{Ge}_{0.30}$/$\mathrm{Si}$/$\mathrm{Si}_{0.70}\mathrm{Ge}_{0.30}$ heterostructure, which is grown on multiple buffer layers on a (001) CVD-grown silicon wafer. All silicon in this experiment is of natural isotope composition. The tensile strained, nominally \SI{10}{\nano\meter} thick $\mathrm{Si}$ layer forms a quantum well hosting a two-dimensional electron gas. The upper $\mathrm{Si}_{0.70}\mathrm{Ge}_{0.30}$ layer has a nominal thickness of \SI{30}{\nano\meter} and is capped with \SI{2}{\nano\meter} of $\mathrm{Si}$. The lower $\mathrm{Si}_{0.70}\mathrm{Ge}_{0.30}$ layer has a nominal thickness of \SI{2}{\micro\meter} with an intermediate, chemically-mechanically polished step after \SI{1}{\micro\meter}. Beneath the buffer, a nominally \SI{3}{\micro\meter} thick linearly graded $\mathrm{Si}_{1-x}\mathrm{Ge}_{x}$ ($x\in[0.05, 0.3]$) buffer is deposited on the substrate silicon wafer.

Ohmic contacts are fabricated by phosphorus ion implantation and thermal activation at \SI{730}{\celsius} for \SI{30}{\second}. The three metal gate layers are made of a $\mathrm{Ti}$/$\mathrm{Pt}$ stack (\SI{5}{\nano\meter} and \SI{15}{\nano\meter}/\SI{22}{\nano\meter}/\SI{29}{\nano\meter}, respectively), shaped by \SI{100}{\kilo\electronvolt} electron-beam lithography (schematics of gate layers in \fixref{ext:fig:extended_data_figure_6}.) and lift-off, isolated from each other and the substrate by amorphous, \SI{10}{\nano\meter} thick layers of $\mathrm{Al}_2\mathrm{O}_3$, fabricated by atomic layer deposition.

\subsection*{Device design and operation}
The lowest gate layer is comprised of three large, grounded screening gates (ScrTL, ScrTR, ScrBR) transversely confining two $\approx$ \SI{5}{\micro\meter} and $\approx$ \SI{10}{\micro\meter} long one-dimensional electron channels (1DECs) joining at a \SI{90}{\degree} angle in the device center, while confinement into quantum dots along the channels is achieved via 4 claviature gates per channel in gate layers two (LS2, RS2, TS2, LS4, RS4 and TS4) and three (LS1, RS1, TS1, LS3, RS3 and TS3).
Six individual gates (JLS, JLP, JP, JRP, JRS and JTS) could enable independent control in the critical junction center.
The electron channels are terminated towards the edges of the device with three individual gates each (TLB1/2, TLP, TRB1/2, TRP, TTB1/2 and TTP), beyond which a single electron transistor (SET) each is formed.
The right SET serves as an electron reservoir and proximal charge detector \cite{klosCalculationTunnelCouplings2018}.
In total 218 claviature gates span the 1DECs, forming the two shuttle lanes.

Due to the periodic gating defining the shuttle lanes, several gates in each lane are grouped into gate sets, following the color coding in \fixref{fig:figure1}.
To all gates within one gate set, the same voltage is applied.
The eight periodic gate sets are: $\mathrm{S}_{\mathrm{x}, 1} =$ TLP, LS3, JRP and RS1; $\mathrm{S}_{\mathrm{x}, 2} =$ TLB2, LS4, JLS, JRS, TRB2 and RS4; $\mathrm{S}_{\mathrm{x}, 3} =$ LS1, JLP and RS3; $\mathrm{S}_{\mathrm{x}, 4} =$ LS2, JP and RS2; $\mathrm{S}_{\mathrm{y}, 1} =$ TTP and TS3; $\mathrm{S}_{\mathrm{y}, 2} =$ TTB2 and TS4; $\mathrm{S}_{\mathrm{y}, 3} =$ TS1; $\mathrm{S}_{\mathrm{y}, 4} =$ TS2 and JTS.
Gates belonging to these gate sets, as well as TRB1, TRP and RB1 are controlled by voltage pulses.

A constant and identical DC voltage of \SI{330}{\milli\volt} is applied to all gates belonging to gate sets $\mathrm{S}_{j, i}$ ($i\in\{1,2,3,4\}$, $j\in\{\mathrm{x}, \mathrm{y}\}$).
TLB1 and TTB1 are grounded during the measurements presented in this work.
TRP, TRB1 and the gates defining the right SET are tuned to individual DC voltages to facilitate charge initialization and readout.

A pulse $P$ is defined as the set of time-dependent voltage sequences simultaneously applied as $\Delta V_g(t)$ individually specified for one or more gates $g$. $P$ is applied to $g$ in addition to the constant DC voltage specific to $g$.
The voltage applied to gates not specified in the definition of $P$ remains constant during the duration of the pulse. The time reverse and $n$ consecutive repetitions of $P$ are denoted as $P^{-1}$ and $P^n$, respectively. Pulses can be composed of other pulses by applying these immediately after each other.

We define the idle state of the device such that a \qd is formed under every fourth gate, $\lambda =$ \SI{280}{\nano\meter} in each shuttle lane by applying a suitable periodic confinement potential.
In this idle configuration, the \qd minima in the x-shuttle are centered under the $\mathrm{S}_{\mathrm{x}, 1}$ gate set and in the y-shuttle under the $\mathrm{S}_{\mathrm{y}, 1}$ gate set (\fixref{fig:figure1}f).
Additionally, in order to place \qdn{X}{0} in closer proximity to the SET at the right end of the x-shuttle for readout, the readout state is defined, placing the \qd minima in the x-shuttle under $\mathrm{TRP}$ and the $\mathrm{S}_{\mathrm{x}, 3}$ gate set. If not otherwise stated, atomic pulses start at and return to the idle state.
We label the \qds by their nominal position as indicated in \fixref{fig:figure1}f.
While conveyor-mode shuttling continuously moves the \qds, we reassign the \qd labels after each atomic pulse to match the initial \qd positions. A simulation of the electron wavefunction when confined in the \qds is shown in \fixref{fig:figure1}g (more details in Supplementary Note I).

Starting and ending in the readout state, the charge initialization atomic pulses (\fixref{fig:figure1}b) are calibrated from charge stability measurements to initialize the charge occupancy of \qdn{X}{0} to either one electron ($\mathrm{L}_1$) or zero ($\mathrm{L}_0$), by transferring electrons between \qdn{X}{0} and a reservoir formed at the SET at the right end of the x-shuttle, similar to \refa\cite{xueSiSiGeQuBus2024}.
The charge occupancy of \qdn{X}{0} in the readout state is determined from the current across the right SET, which is tuned to its optimal sensitivity by applying the atomic pulse $\mathrm{R}_\mathrm{CC}$.
To reset the \qd occupancy for future readout, $\mathrm{R}_\mathrm{CC}$ is always followed by unloading \qdn{X}{0}, defining the destructive readout pulse $\mathrm{R}$.

From $\lambda_\mathrm{x}$, $\lambda_{\mathrm{L\rightarrow R}}$ and $\lambda_{\mathrm{R\rightarrow L}} = \lambda^{-1}_{\mathrm{L\rightarrow R}}$ are derived to transition between the idle and readout state (\cref{fig:figure1}c).

All finally composed pulses are either led or followed by a $\mathrm{ROI}$ pulse (\fixref{fig:figure1}e), used to recalibrate the SET current levels during the readout process. During each readout (i.e. application of $\mathrm{R}_\mathrm{CC}$), the current through the SET is monitored and averaged in the time interval $t_\mathrm{RO}$.
Due to the capacitive coupling of the charge present in \qdn{X}{0} in the readout state, the chemical potential of the SET's sensing is altered, leading to proximal charge detection by the current across the SET.

We can use the resulting current to readout the number of electrons occupying \qdn{X}{0} using the following procedure:
First, the current background as well as current thresholds allowing determination of the number of electrons in \qdn{X}{0} are extracted.
The background and thresholds are calculated block-wise over differing numbers of repetitions to accommodate changing system noise and SET current drift.
The background in one pulse repetition is calculated by averaging the measured current during all readouts in each pulse repetition and then averaged over the current block of repetitions.
The background is subtracted from the measured current.
The reference readouts directly following the $\mathrm{L}_1$ and $\mathrm{L}_0$ atomic pulses within the ROI pulse contained in each pulse repetition are used to determine the current thresholds discriminating the number of electrons in \qdn{X}{0}.
To make the observed threshold robust against unintentionally incorrect charge occupancy of \qdn{X}{0}, we remove the \SI{5}{\percent} largest and smallest readout currents each from the calculation.
The threshold discriminating between zero and one electrons is set as the mid-point between the average current in the reference readouts and then averaged over the current block of repetitions.

It becomes apparent that in $\approx0.04\,$\si{\percent} of readouts, two electrons are unexpectedly detected in \qdn{X}{0}.
While a clear distinction between outliers in the current during the detection of one electron and the detection of two electrons can not always be made, we still attempt to make this discrimination.
The corresponding threshold discriminating between one and two electrons detected is set $4\cdot\sigma_{\mathrm{L}_1}$ beyond the average current during the reference readout after $\mathrm{L}_1$, where $\sigma_{\mathrm{L}_1}$ is the population standard deviation of the average measured currents during this readout, taken over the current block of repetitions.

To clear the otherwise fully depleted shuttle lanes from remaining charges of previous experiments, a flush pulse, similar to \refa\cite{xueSiSiGeQuBus2024} (\fixref{ext:fig:extended_data_figure_5}a), is applied at irregular intervals using $A_\mathrm{x}=A_\mathrm{y}=A_\mathrm{J} = $ \SI{260}{\milli\volt}.

\subsection*{Experimental setup}
The device is mounted on the mixing chamber stage of an Oxford Instruments wet dilution refrigerator at a temperature of $\approx$ \SI{30}{\milli\kelvin}.
Signals inside the refrigerator are carried to the sample via custom-made high density flexible flat cables with a bandwidth on the order of \SI{1}{\mega\hertz}.
This bandwidth allows for experimentation close to the frequency range used for spin-coherent single electron shuttling \cite{langrockBlueprintScalableSpin2023}.
At room temperature, voltages applied to gates that are held constant are generally applied using custom-made low noise digital to analog converters.
Voltages are applied to pulsed gates using Zürich Instruments HDAWG arbitrary waveform generators.
The two ohmic contacts of the SET used for readout are connected to room-temperature bias-tees allowing for an offset voltage as well as a modulated voltage to be applied and a modulated current to be measured.
For the results presented here, the offset voltage is set to \SI{0}{\volt} while the modulated voltage on the order of \SI{100}{\micro\volt} is applied using the output of a Zürich Instruments MFLI lock-in amplifier.
The modulated current is then converted to a voltage by a Basel Precision Instruments transimpedance amplifier which is measured using the lock-in amplifier.
For ease of integration, the demodulated signal is converted to an appropriately scaled voltage output by the lock-in amplifier, buffered by a Stanford Research Systems SR560 preamplifier and acquired using an AlazarTech ATS9440 waveform digitizer.

\subsection*{Fidelity calculation and error estimation}
We consider the result of a shuttle pulse containing several atomic readout pulses a success, if electrons are detected during and only during the readout pulses for which we expect electrons to be detected.
Thus, a pulse in which an electron is detected where no electron is expected is considered a failed pulse.
If $n_{\mathrm{s}}$ is the number of successful pulse repetitions and $n_{\mathrm{total}}$ is the total number of repetitions, the pulse fidelity $F_\mathrm{total}$ is defined as:
\begin{gather*}
    F_\mathrm{total} =  \frac{n_{\mathrm{s}}}{n_{\mathrm{total}}}.
\end{gather*}
The generally asymmetric uncertainty on $F_\mathrm{total}$ is determined from the two-sided, \SI{95}{\percent} Clopper-Pearson confidence interval $\left[F^-_{\mathrm{total},\ 95\,\%}, F^+_{\mathrm{total},\ 95\,\%}\right]$ as $\sigma^\pm_{F_\mathrm{total}} = F^\pm_{\mathrm{total},\ 95\,\%} - F_\mathrm{total}$.
The value of $F_\mathrm{total}$ and its uncertainty is denoted as \fidelitywop{(F_\mathrm{total})}{\sigma^+_{F_\mathrm{total}}}{\sigma^-_{F_\mathrm{total}}}.

Occasionally, we observe from the SET current level that two electrons instead of one expected electron are detected. For these rare cases, we still label the repetition as successful, as we attribute the observation to the unintended loading of an additional electron presumably during the intended electron loading, which constitutes a SPAM error.
The detection of an additional electron in the \qd, in which one electron is expected due to the application of $\mathrm{L}_1$, is explained by the conveyor-mode shuttle operating with doubly occupied \qds.
While the determination of this observation as a success reduces the impact of SPAM errors on the pulse fidelity, it is stricter on the fidelity of charge shuttling, as the confinement of two electrons in a \qd is expected to be weaker and therefore more prone to shuttle errors than shuttling only one electron, e.g. one electron tunneling into a neighboring, unoccupied \qd.

\subsection*{Fidelity calculation for looping in the T-junction}
The fidelity per charge loop $F_1$ and the corresponding uncertainties $\sigma^\pm_{F_1}$ are calculated following:
\begin{gather*}
    F_1 = F_\mathrm{total}^{\frac{1}{n_\mathrm{loops}}} \\
    \sigma^\pm_{F_1} = (F^\pm_{\mathrm{total},\ 95\,\%})^{\frac{1}{n_\mathrm{loops}}} - F_1
\end{gather*}
These values are denoted as $(F_1)^{\sigma^+_{F_1}}_{\sigma^-_{F_1}}$.

\section*{Data Availability}
The data generated in this study are available in the Zenodo repository under https://doi.org/10.5281/zenodo.18032292.

\section*{Acknowledgements}
We acknowledge the support of the Dresden High Magnetic Field Laboratory (HLD) at the Helmholtz-Zentrum Dresden - Rossendorf (HZDR), member of the European Magnetic Field Laboratory (EMFL). This work was funded by the German Research Foundation (DFG) within the project 421769186 (SCHR 1404/5-2) and under Germany's Excellence Strategy - Cluster of Excellence Matter and Light for Quantum Computing (ML4Q) EXC 2004/2 - 390534769, and by the European Union’s Horizon Research and Innovation Actions under Grant Agreement No. 101174557 (QLSI2).
The device fabrication has been done at HNF - Helmholtz Nano Facility, Research Center Juelich GmbH \cite{albrechtHNFHelmholtzNano2017}.

\section*{Author contributions statement}
M.B. carried out the experiments and simulation presented in this work. M.B. analyzed the data with L.R.S. Sample screening and pre-characterization was carried out by L.D. guided by M.B.. Custom components and the control electronic scheme were designed by M.B.. M.B. adapted the device fabrication process developed by R.X., J.S.T., and I.S. and carried out the fabrication. Electron beam lithography was carried out by S.T.. M.B. and P.S. operated the dilution refrigerator. L.R.S. and M.B. designed the device and L.R.S. supervised the study. L.R.S. and H.B. provided guidance to all authors. M.B. and L.R.S wrote the manuscript, which was commented by all other authors.

\section*{Competing interests}
H.B, L.R.S., R.X., I.S. are co-inventors of patent applications that cover conveyor-mode shuttling and/or its applications. L.R.S. and H.B. are founders and shareholders of ARQUE Systems GmbH. The other authors declare no competing interest.

\clearpage
\begin{table}[h!]
    \centering
    \caption{Shuttle fidelities at $A_{\mathrm{x}} = A_{\mathrm{y}} = A_{\mathrm{J}} = 260\,\mathrm{mV}$ and  $v_\lambda = v_\mathrm{J} = $ \SI{270}{\milli\meter\per\second}. Shuttle paths are indicated as sequences of key \qds traversed in time-order. $\mathrm{X}_{0}\left(\mathrm{X}_{16}\mathrm{Y}_{1}\mathrm{X}_{19}\right)^{10^6}\mathrm{X}_{0}$ with $A_{\mathrm{J}} = $ \SI{100}{\milli\volt} yields $F_1 = $ \fidelity{100.0000000}{+0}{-9\times 10^{-7}}.}
    \label{tab:table1}
    \renewcommand{\arraystretch}{1.2}
    \setlength{\tabcolsep}{5pt}
    \begin{tabularx}{\columnwidth}{>{\raggedright\arraybackslash}X >{\raggedright\arraybackslash}p{1.4cm} >{\raggedright\arraybackslash}p{3cm}}
        \toprule
            shuttle path & {$F_\mathrm{total}$ (\%)} & {$F_1$ (\%)} \\
        \midrule
            $\mathrm{X}_{0}\mathrm{X}_{4}\mathrm{X}_{0}$ & $99.9^{+0.1}_{-0.5}$ \\
            $\mathrm{X}_{0}\mathrm{X}_{35}\mathrm{X}_{0}$ & $100.0^{+0.0}_{-0.4}$ \\
            $\mathrm{X}_{0}\mathrm{X}_{17}\mathrm{Y}_{0}\mathrm{Y}_{4}\mathrm{Y}_{0}\mathrm{X}_{17}\mathrm{X}_{0}$ & $100.0^{+0.0}_{-0.4}$ \\
            $\mathrm{X}_{0}\mathrm{X}_{17}\mathrm{Y}_{0}\mathrm{Y}_{17}\mathrm{Y}_{0}\mathrm{X}_{17}\mathrm{X}_{0}$ & $100.0^{+0.0}_{-0.4}$ \\
            $\mathrm{X}_{0}\left(\mathrm{X}_{16}\mathrm{Y}_{1}\mathrm{X}_{19}\right)^{10^5}\mathrm{X}_{0}$ & $100.0^{+0.0}_{-0.4}$ & \fidelitywop{100.00000000}{+0}{-4\times 10^{-6}} \\
            $\mathrm{X}_{0}\left(\mathrm{X}_{16}\mathrm{Y}_{1}\mathrm{X}_{19}\right)^{10^6}\mathrm{X}_{0}$ & $99.8^{+0.2}_{-0.9}$ & \fidelitywop{99.9999998}{+2\times 10^{-7}}{-9\times 10^{-7}} \\
        \bottomrule
    \end{tabularx}
\end{table}

\clearpage
\onecolumngrid
\section*{Extended data}
\setcounter{figure}{0} 
\renewcommand\figurename{Extended Data Fig.}
\FloatBarrier

\begin{figure*}
    \centering
    \includegraphics[scale=1]{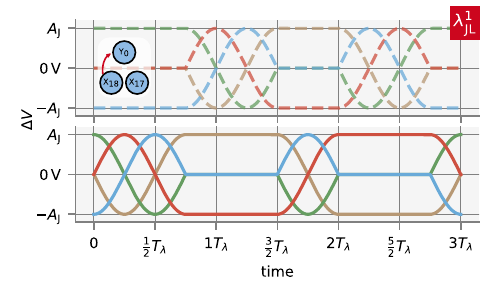}
    \caption{\textbf{Junction transfer atomic pulse $\lambda^1_\mathrm{JL}$} This figure follows the legend in \fixref{fig:figure1}. The atomic pulse shown shuttles the charge in \qdn{X}{18} to \qdn{Y}{0}.}
    \label[extfigure]{ext:fig:extended_data_figure_1}
\end{figure*}

\begin{figure*}
    \centering
    \includegraphics[scale=1]{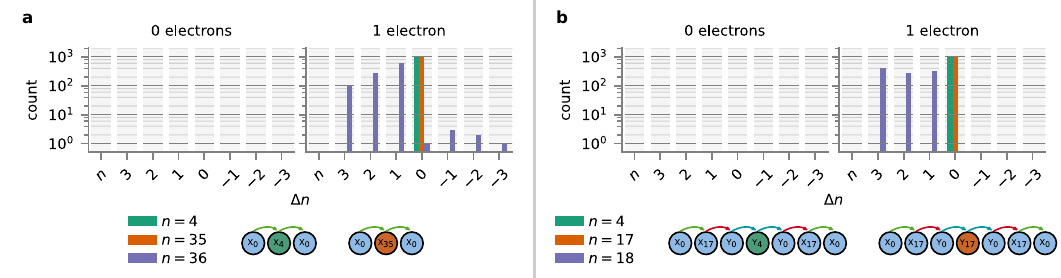}
    \caption{\textbf{Verifying shuttling in the x- and y-shuttles.} Complementary results of \fixref{fig:figure2} at $v_\lambda = v_\mathrm{J} =$ \SI{28}{\milli\meter\per\second}. \textbf{a} Shuttling in the x-shuttle. \textbf{b} Shuttling in the x-shuttle up to \qdn{X}{17} followed by the y-shuttle.}
    \label[extfigure]{ext:fig:extended_data_figure_2}
\end{figure*}

\begin{figure*}
    \centering
    \includegraphics[scale=1]{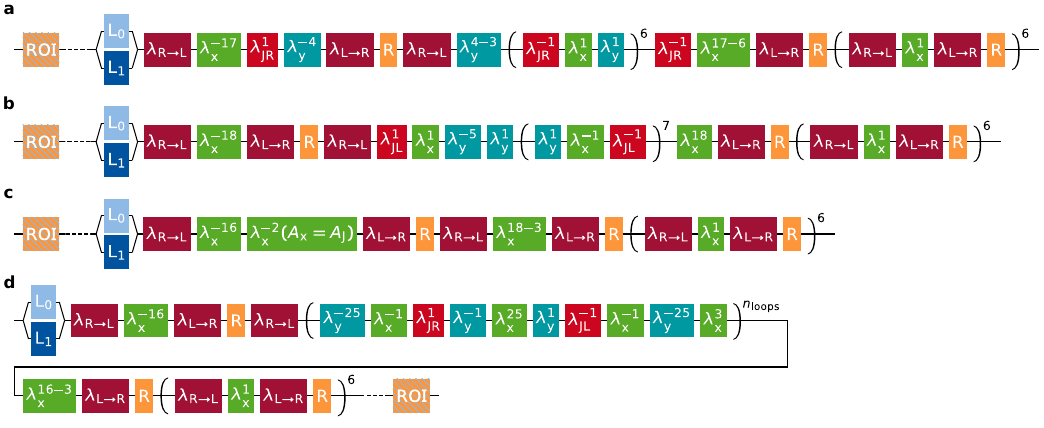}
    \caption{\textbf{Pulses applied to explore parameter space of junction shuttling.} \textbf{a}-\textbf{c} Pulses used for results in \fixref{fig:figure3}a. \textbf{d} Pulse used for results in \fixref{fig:figure3}b and \fixref{fig:figure3}c. Note that pairs of $\lambda_\mathrm{L \rightarrow R}\lambda_\mathrm{R \rightarrow L}$ and their reverse are removed by a simplification step before playback.}
    \label[extfigure]{ext:fig:extended_data_figure_3}
\end{figure*}

\begin{figure*}
    \centering
    \includegraphics[scale=1]{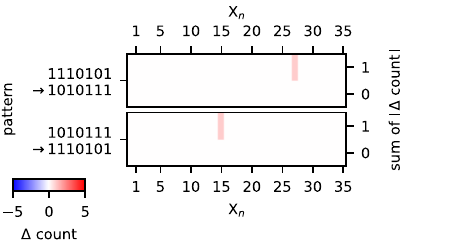}
    \caption{\textbf{Error counts for pattern SWAP.} Complementary data for \fixref{fig:figure4}d following the same color map.}
    \label[extfigure]{ext:fig:extended_data_figure_4}
\end{figure*}

\begin{figure*}
    \centering
    \includegraphics[scale=1]{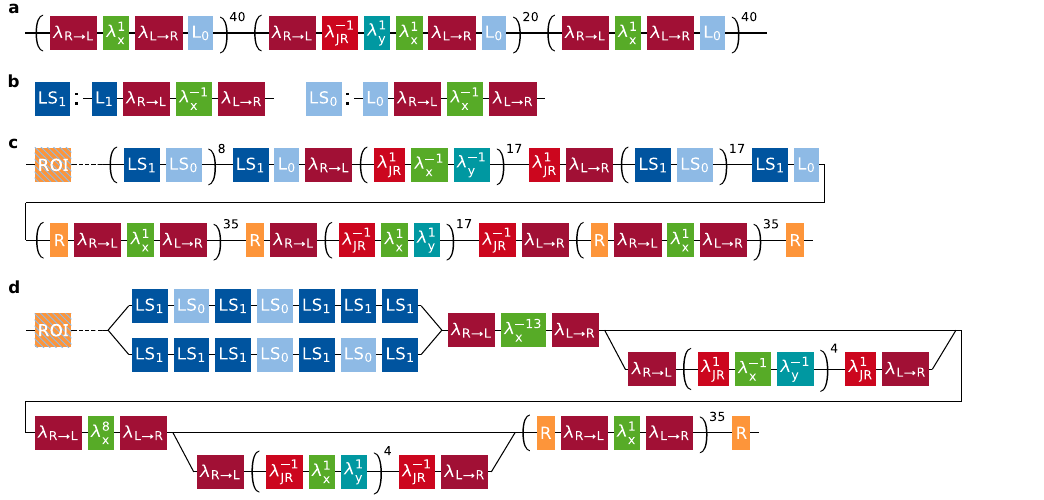}
    \caption{\textbf{Additional pulses.}  \textbf{a} Flush pulse. Due to the extensive use of slow $\mathrm{L}_0$ pulses, the flush is not performed before every pulse repetition. During the flush $A_\mathrm{J} = $ \SI{260}{\milli\volt} and the same shuttle velocities used during the rest of the pulse are applied. \textbf{b} Load and shuttle notation shorthand. \textbf{c} Exemplary pulse for shuttle of $\overline{10}$-electron pattern. For other electron patterns, only the initialization pulses are modified accordingly. \textbf{d} Pulse applied in \fixref{fig:figure4}c and \fixref{fig:figure4}d to perform SWAP of an electron pattern. The top and bottom initialization branches (first row) as well as the top and bottom shuttling branches (second row) are combined to yield 4 different pulses. The bottom shuttling branch performs the SWAP. Note that pairs of $\lambda_\mathrm{L \rightarrow R}\lambda_\mathrm{R \rightarrow L}$ and their reverse are removed by a simplification step before playback.}
    \label[extfigure]{ext:fig:extended_data_figure_5}
\end{figure*}

\begin{figure*}
    \centering
    \includegraphics[scale=1]{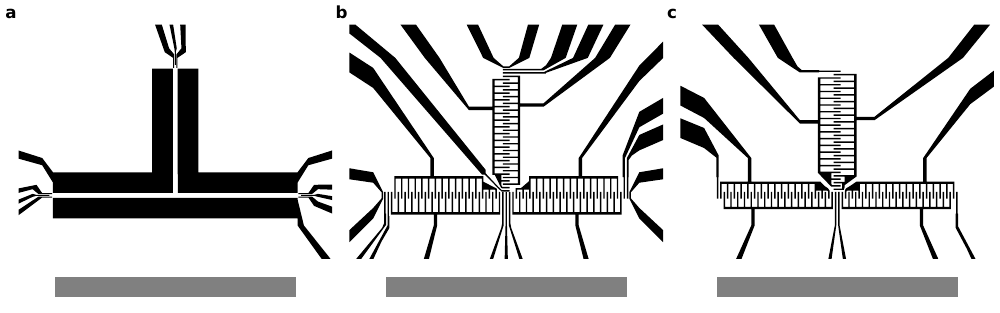}
    \caption{\textbf{Metal gate layers defining the device.} Dark shapes correspond to metalized regions visible in \fixref{fig:figure1}a. Scale bars are \SI{10}{\micro\meter}. \textbf{a} First metal layer. \textbf{b} Second metal layer. \textbf{c} Third metal layer.}
    \label[extfigure]{ext:fig:extended_data_figure_6}
\end{figure*}

\newpage
\FloatBarrier
\section*{Supplementary Information}

\onecolumngrid
\setcounter{figure}{0} 
\renewcommand\figurename{Supplementary Figure}

\subsection*{Supplementary Note I: Simulation of shuttling in the T-junction} 
\FloatBarrier
We simulate the wavefunction of an electron confined to a quantum dot (\qd) in the device by solving the one-electron time independent Schroedinger equation on a two-dimensional grid on the gate-controlled instantaneous potential landscape at many time steps within a suitable region around the expected electron position.
The total potential within the Si/SiGe quantum well for each time step is calculated in COMSOL Multiphysics\textsuperscript\textregistered\xspace using a model of the device.
During the calculation we disregard charge screening within the quantum well as the relevant device region is expected to be almost fully depleted.
To verify continuous electron motion, we evaluate the expectation value of the ground state position $\langle x_i\rangle = \braketm{g}{x_i}{g}$, where $x_i\in\{\mathrm{x}, \mathrm{y}\}$ is the x- or y-coordinate, and $\ket{\mathrm{g}}$ is the electron ground state.
Analogous to \refa\cite{sup:langrockBlueprintScalableSpin2023}, we also calculate the resulting orbital splitting $\mathrm{E}_\mathrm{orb} = \braketm{\mathrm{e}}{H}{\mathrm{e}} - \braketm{\mathrm{g}}{H}{\mathrm{g}}$ between $\ket{\mathrm{g}}$ and the first  excited state $\ket{\mathrm{e}}$, where $H$ is the system Hamiltonian.

\begin{figure}
    \centering
    \includegraphics[scale=1]{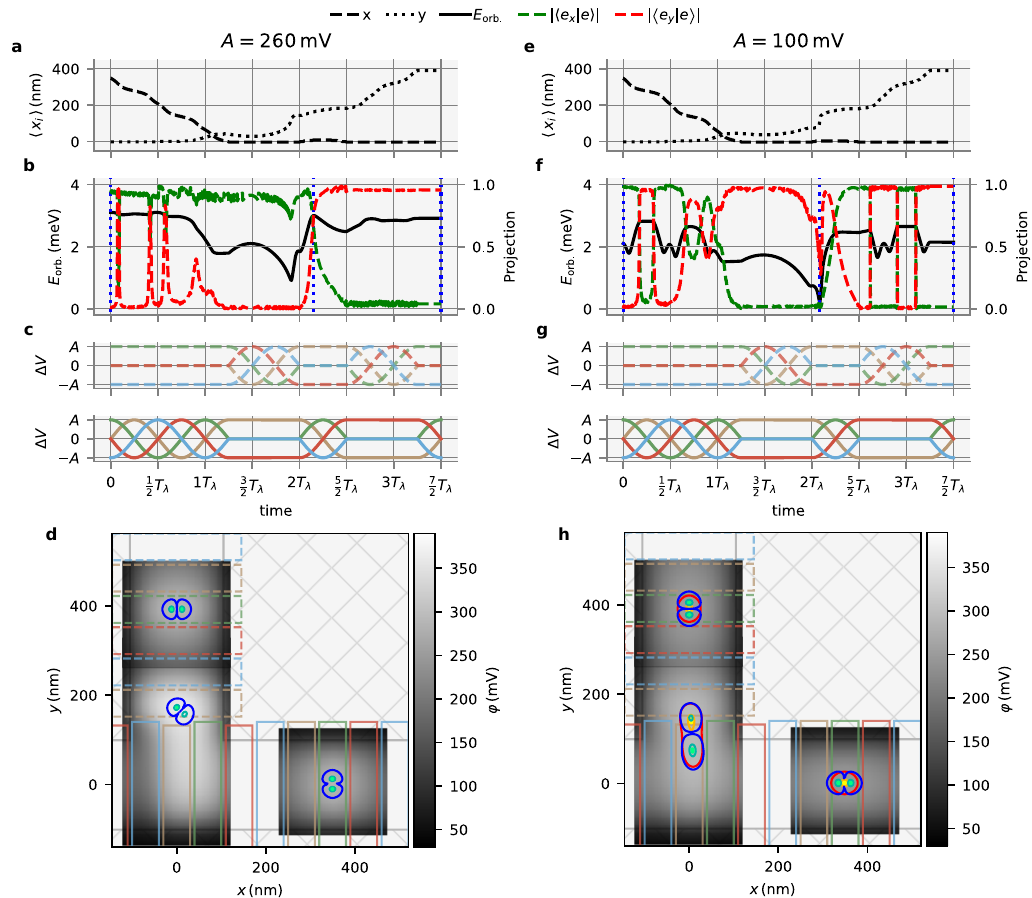}
    \caption{\textbf{Additional simulation results for $A= $ \SI{260}{\milli\volt} and $A= $ \SI{100}{\milli\volt}.} $A_\mathrm{x} = A_\mathrm{y}= A_\mathrm{J} = A$. \textbf{a, e} Expectation value of electron x and y position in the ground state.  \textbf{b, f} Orbital splitting and wavefunction projection during transfer. Blue markers indicate the time steps shown in (\textbf{d, h}), respectively. \textbf{c, g} Corresponding applied voltage pulses. Colors follow the gate color-coding in \fixref{fig:figure1}. \textbf{d} First excited states for the time steps shown in (\textbf{b}), corresponding to time steps shown in \fixref{fig:figure1}g. Teal, green and blue sequence correspond to the $95\,\%$, $68\,\%$, $1\,\%$ levels of the probability density for the first excited state. \textbf{h} Ground states and first excited states for the time steps shown in (\textbf{f}). \SI{95}{\percent}, \SI{68}{\percent}, \SI{1}{\percent} probability density levels of the ground states shown in yellow, orange and red, respectively. First excited states follow color map in (\textbf{d}).}
    \label[extfigure]{sup:fig:supplementary_figure_1}
\end{figure}

As the central device aspect, we simulate shuttling from \qdn{X}{16} to \qdn{Y}{0} using $A_\mathrm{x} = A_\mathrm{y} = A_\mathrm{J} = A$, twice using $A = $ \SI{260}{\milli\volt} and $A = $ \SI{100}{\milli\volt}, respectively as well as similar DC voltages as used in the experiment. 
The results are shown in \fixref{sup:fig:supplementary_figure_1}a-c and \fixref{sup:fig:supplementary_figure_1}e-g, with probability densities for key time steps rendered in \fixref{sup:fig:supplementary_figure_1}d and \fixref{sup:fig:supplementary_figure_1}h.
Renderings of full simulations are shown in \fixref{sup:vid:video1} ($A = $ \SI{260}{\milli\volt}) and \fixref{sup:vid:video2} ($A = $ \SI{100}{\milli\volt}).

Coinciding with the respective minimum in $\mathrm{E}_{\mathrm{orb}}$ for both drive amplitudes, a fast increase in $\langle y\rangle$ is obtained.
This increase is more pronounced for $A = $ \SI{100}{\milli\volt} compared to $A = $ \SI{260}{\milli\volt}, however both $\langle x\rangle$ and $\langle y\rangle$ are continuous for both drive amplitudes indicating that electron transfer occurs without long-range tunneling (\fixref{sup:fig:supplementary_figure_1}a, e).

The minima in $\mathrm{E}_{\mathrm{orb}}$ occur near $2T_\lambda$, decreasing to $\mathrm{E}_{\mathrm{orb}, \mathrm{min}} \approx 910\,\mathrm{\mu eV}$ and $\mathrm{E}_{\mathrm{orb}, \mathrm{min}} \approx 185\,\mathrm{\mu eV}$ for $A = $ \SI{260}{\milli\volt} and $A = $ \SI{100}{\milli\volt}, respectively.
Further inspection reveals the cause for this decrease as the lowered confinement when merging the \qd containing the electron and the first \qd of the vertical conveyor.
As $\mathrm{E}_{\mathrm{orb}}$ decreases, excitation into the excited state may occur, even at low shuttle velocities.
If the occupancy of these states is uncontrolled and orbital relaxation occurs slowly or excitations occur often, spin state decoherence occurs, as the electron $g$-factor and valley splitting are expected to differ significantly between ground and excited states \cite{sup:volmerMappingValleySplitting2024,sup:struckSpinEPRpairSeparationConveyormode2024}, warranting further investigation.
We estimate that realistic potential disorder, caused by e.g. charged defects in the semiconductor-oxide interface above the Si capping layer, could decrease $\mathrm{E}_{\mathrm{orb}, \mathrm{min}}$ into a problematic regime for $A = $ \SI{100}{\milli\volt} (c.f. \refa \cite{sup:langrockBlueprintScalableSpin2023}).

When sufficiently far away from the junction center, a consistent primary confinement axis with respect to the shuttle direction is observed, when the electron is centered below a gate.
For $A = $ \SI{260}{\milli\volt}, this axis is aligned longitudinally with the shuttle direction (indicating stronger confinement by shuttle gates) and for $A = $ \SI{100}{\milli\volt} it is aligned transversely (indicating stronger confinement by screening gates).
We refer to the initial and final excited states during the simulation as $\ket{\mathrm{e}_{\mathrm{x}}}$ and $\ket{\mathrm{e}_{\mathrm{y}}}$, respectively, and plot the projection of the instantaneous first excited state onto these states in \fixref{sup:fig:supplementary_figure_1}b and \fixref{sup:fig:supplementary_figure_1}f.

For $A = $ \SI{260}{\milli\volt}, the first excited states remain well aligned with the primary confinement axes, except for short lived rotations when the electron is centered between two gates.
Using this metric, the moment of transfer between the x- and y-shuttle can be identified when the first excited state rotates by \SI{90}{\degree}, coinciding with a maximum in $\mathrm{E}_{\mathrm{orb}}$.

For $A = $ \SI{100}{\milli\volt}, this rotation still occurs during this transfer, however similar rotations occur at several time steps as the confinement strength varies with the height of the gate the electron is currently under.
Tuning drive amplitude and DC voltage separately for each gate layer is expected to suppress these additional rotations.
Additionally, no maximum in $\mathrm{E}_{\mathrm{orb}}$ is observed in the moment of transfer due to the weak confinement in the junction center, as indicated by the low $\mathrm{E}_{\mathrm{orb}, \mathrm{min}}$ and the non-circular electron probability density.

\end{document}